%% file: 2arxiv.tex
\documentclass[preprint,12pt,authoryear]{elsarticle}

\usepackage{amssymb}
 \usepackage{amsthm}

\usepackage{amsmath}

\usepackage{pstricks}
\usepackage{pstricks-add}
\usepackage{pst-node}
\usepackage{pst-plot}
\SpecialCoor

\usepackage[super]{nth}

\usepackage{algorithm}
\usepackage{algpseudocode}

\usepackage{enumitem}

\newtheorem{theorem}{Theorem}

\journal{Theoretical Computer Science}

\begin{document}

\begin{frontmatter}

\title{Small Longest Tandem Scattered Subsequences} 

\author[ist]{Lu\'is~M.~S.~Russo \corref{cor1}}
\ead{luis.russo@tecnico.ulisboa.pt}
\address[ist]{INESC-ID and the Department of Computer Science and Engineering,
  Instituto Superior T\'ecnico, Universidade de Lisboa.
}
\cortext[cor1]{Corresponding author.}

\author[ist]{Alexandre~P.~Francisco}
\ead{aplf@tecnico.ulisboa.pt}

\begin{abstract} 
  We consider the problem of identifying tandem scattered subsequences
  within a string. Our algorithm identifies a longest subsequence which
  occurs twice without overlap in a string. This algorithm is based on the
  Hunt-Szymanski algorithm, therefore its performance improves if the
  string is not self similar.  This occurs naturally on strings over large
  alphabets. Our algorithm relies on new results for data structures that
  support dynamic longest increasing sub-sequences. In the process we also
  obtain improved algorithms for the decremental string comparison problem.

\end{abstract}


\end{frontmatter}

\section{Introduction} 
\label{sec-Introduction}
In this paper we study longest common scattered sub-sequences (LCSS). Given
two strings $P$ and $S$ the LCSS is used extensively as a measure of
similarity. In particular we consider a variant of this problem, where the
LCSS must occur twice without overlap in an initial string $F$. We study
algorithms and data structures that are relevant for this goal. Namely we
use the Hunt-Szymanski algorithm~\citeyearpar{hunt1977fast} and
present new results for data structures that maintain information about the
longest increasing sub-sequence of a dynamic sequence of numbers and new
algorithms for the decremental string comparison problem. Specifically we
obtain the following results:
\begin{enumerate}
\item A data structure to maintain the longest increasing subsequence (LIS)
  of a dynamic list of numbers. This structure can be used to efficiently:
  \texttt{Append} a number at the end of the list; remove the current
  minimum value from the sequence (\texttt{ExtractMin}); obtain a current
  longest increasing subsequence (\texttt{GetLIS}). When the list contains
  $\ell \geq 2$ elements, the operation \texttt{Append} requires
  $O(\log \ell)$ time\footnote{Note that to simplify expressions as
    $O(1 + \log \ell)$ we impose restrictions on parameters such as
    $\ell \geq 2$. This also avoids invalid expressions such as when
    $\ell=0$. In general the complexity of the excluded cases is
    $O(1)$.}. If the size of the LIS is $\lambda \geq 2$ then the
  \texttt{ExtractMin} operation requires $O(\lambda \log \ell)$ time.  See
  Theorem~\ref{teo:BasicLIS}. We further improve these bounds by analyzing
  batches of operations and assuming the final sequence is empty. In this
  case \texttt{ExtractMin} requires
  $O(\lambda (1 + \log(\min\{\lambda, \ell/\lambda\})))$ amortized time per
  operation and \texttt{Append} requires $O(1)$ amortized time, when the
  numbers are inserted in decreasing sequences of size at least $\lambda$
  elements. See Theorem~\ref{teo:LISA}. This structure uses optimal
  $O(\ell)$ space.

\item Using the~\citet{hunt1977fast} reduction from LCSS to LIS we obtain
  new bounds for the decremental string comparison problem. In particular,
  for a given string $F$ of size $n>1$, we show that it is possible to
  obtain all the LCSS values for all the pairs of strings $P$ and $S$ such
  that $F=P.S$ in
  $O(\min\{n,\ell\} \lambda (1 + \log (\min\{\lambda, \ell/\lambda\})) + n
  \lambda + \ell)$ time, where $\lambda \geq 2$ is the size of the LCSS and
  $\ell \geq 2$ is the number of pairs of positions in $F$ that contain the
  same letter. Therefore it is possible to determine the LTSS within this
  time, i.e., the LCSS which occurs twice without overlap in a string $F$.
\end{enumerate}

\section{The problem} 
\label{sec-The_problem}
Let us start by describing the longest tandem scattered sub-sequence (LTSS)
problem of a given string $F$. We will use a running example with
$F = \mathtt{AGCGAACGGGTA}$. The meaning of tandem is that the
sub-sequence needs to occur twice without overlap in $F$. Therefore $F$ can
be partitioned into a prefix $P$ and a suffix $S$, i.e., $F=P.S$, such that
the desired scattered sub-sequence is a longest common scattered
sub-sequence (LCSS) between $P$ and $S$. To determine which partition
yields the overall largest sub-sequence it is necessary to test all such
partitions.

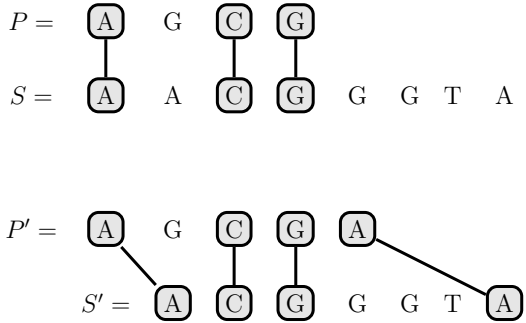
\begin{figure}[htb]
  \begin{center}
    \begin{pspicture}[showgrid=false](0,-0.5)(7,4)
      \psset{linewidth=0.05,framearc=0.6,fillcolor=black!10,fillstyle=solid}
      \begin{Scalebox}{0.80}
      \begin{psmatrix}[colsep=0.4,rowsep=0.6]
        $P = $&\psframebox{A}&G&\psframebox{C}&\psframebox{G}\\
        $S = $&\psframebox{A}&A&\psframebox{C}&\psframebox{G}&G&G&T&A\\
\\
        $P' = $&\psframebox{A}&G&\psframebox{C}&\psframebox{G}&\psframebox{A}\\
        & $S'= $&\psframebox{A}&\psframebox{C}&\psframebox{G}&G&G&T&\psframebox{A}\\
      \end{psmatrix}
      \ncline{-}{1,2}{2,2}
      \ncline{-}{1,4}{2,4}
      \ncline{-}{1,5}{2,5}

      \ncline{-}{4,2}{5,3}
      \ncline{-}{4,4}{5,4}
      \ncline{-}{4,5}{5,5}
      \ncline{-}{4,6}{5,9}

      \end{Scalebox}
    \end{pspicture}
  \end{center}
  \caption{Example of LCSS for two prefix suffix decompositions of $F$,
    $F=P.S$ and $F=P'.S'$ }
  \label{fig:LCSSex}
\end{figure}

Let us consider the partition with $P =\mathtt{AGCG}$ and
$S = \mathtt{AACGGGTA}$. The LCSS is the longest string that occurs as
a scattered sub-sequence of $P$ and $S$. Figure~\ref{fig:LCSSex}
illustrates that the string \texttt{ACG} is a longest common scatted
sub-sequence of $P$ and $S$. A common sub-sequence can be defined as a set
of pairs $(i,j)$ where $i$ is an index over $P$ and $j$ an index over $S$
and the $i$-th letter of $P$ is equal to the $j$-th letter of $S$,
represented as $P(i) = S(j)$.  All numbers $i$ must be distinct among
themselves and all numbers $j$ must be distinct among themselves. Moreover
sorting the pairs by $i$ must also yield a sorted sequence by $j$. In our
example the LCSS for $P$ and $S$ can be represented by the set
$\{(1,1),(3,3),(4,4)\}$.

Figure~\ref{fig:LCSSex} also shows an LCSS for a second prefix suffix
decomposition of $F=P'.S'$. This second decomposition is related to the
first as $P'=P.\mathtt{A}$ and in fact the LCSS is similar to the previous
LCSS, with the extra character \texttt{A}, i.e., \texttt{ACGA}.

In this example the LCSS between $P'$ and $S'$ is the desired overall LTSS.

\section{Reduction to decremental string comparison} 
\label{sec-Idea}

In this section we present the main ideas of an algorithm that computes
LTSS. Given a string $F$ we can reduce this problem to computing the size
of the LCSSs for all prefix and suffix decompositions of $F$, i.e., for all
$P.S=F$, where $P$ is a prefix and $S$ is a suffix. The resulting tandem
can be obtained from the overall largest LCSS.

This process involves $n$ LCSS computations, when the size of $F$ is
$n$. Each LCSS can be determined with the classical dynamic programming
table between $P$ and $S$. Table $D$ is a bi-dimensional array that stores
integers. Each value $D[i,j]$ represents the size of the LCSS between the
prefix of $P$ with $i$ letters and the prefix of $S$ with $j$ letters. The
coordinate $i$ ranges from $0$ to the size of $P$, likewise coordinate $j$
ranges between $0$ and the size of $S$. The value $0$ represents the empty
prefix.

The values $D[i,j]$ can be computed locally according to the
equalities bellow, where $P(i)$ denotes the $i$-th letter of $P$ and $S(j)$
the $j$-th letter of $S$:
\begin{align}
  D[i,j] &= 0 && \mbox{ if $i = 0$ or $j=0$} \label{c:basis} \\
  D[i,j] &=D[i-1,j-1]+1  &&\mbox{ if $i,j > 0$ and $P(i) = S(j)$} \label{c:equal} \\
  D[i,j] &=\max\{D[i,j-1], D[i-1,j]\} &&\mbox{ if $i,j > 0$ and $P(i) \neq S(j)$} \label{c:diff}
\end{align}

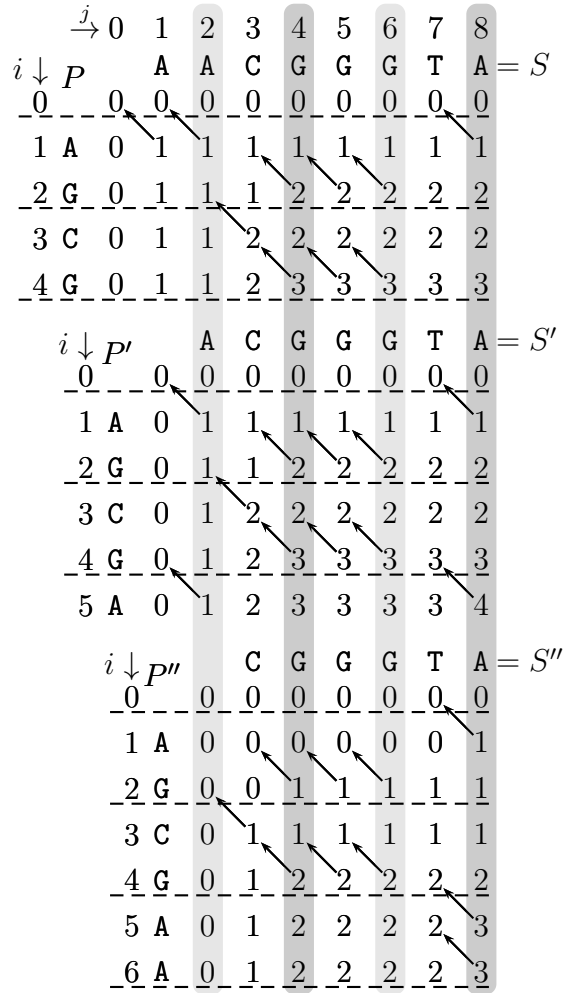
\begin{figure}[tb]
  \begin{center}
    \begin{pspicture}[showgrid=false](6,12)
      \rput[bl](0,0){
        \input{DaDpaDpp.tex}
     }
      \psframe*[linecolor=black!10,framearc=0.7]([nodesep=-0.3,offset=1.3]D_0_2)
      ([nodesep=0.1,offset=-0.3]Dpp_6_0)
      \psframe*[linecolor=black!20,framearc=0.7]([nodesep=-0.3,offset=1.3]D_0_4)
      ([nodesep=0.1,offset=-0.3]Dpp_6_2)
      \psframe*[linecolor=black!10,framearc=0.7]([nodesep=-0.3,offset=1.3]D_0_6)
      ([nodesep=0.1,offset=-0.3]Dpp_6_4)
      \psframe*[linecolor=black!20,framearc=0.7]([nodesep=-0.3,offset=1.3]D_0_8)
      ([nodesep=0.1,offset=-0.3]Dpp_6_6)
        \psline[linestyle=dashed]([offset=-0.2,nodesep=-1.4]D_0_0)([offset=-0.2]D_0_8)
        \psline[linestyle=dashed]([offset=-0.2,nodesep=-1.4]D_2_0)([offset=-0.2]D_2_8)
        \psline[linestyle=dashed]([offset=-0.2,nodesep=-1.4]D_4_0)([offset=-0.2]D_4_8)
        \psline[linestyle=dashed]([offset=-0.2,nodesep=-1.4]Dp_0_0)([offset=-0.2]Dp_0_7)
        \psline[linestyle=dashed]([offset=-0.2,nodesep=-1.4]Dp_2_0)([offset=-0.2]Dp_2_7)
        \psline[linestyle=dashed]([offset=-0.2,nodesep=-1.4]Dp_4_0)([offset=-0.2]Dp_4_7)
        \psline[linestyle=dashed]([offset=-0.2,nodesep=-1.4]Dpp_0_0)([offset=-0.2]Dpp_0_6)
        \psline[linestyle=dashed]([offset=-0.2,nodesep=-1.4]Dpp_2_0)([offset=-0.2]Dpp_2_6)
        \psline[linestyle=dashed]([offset=-0.2,nodesep=-1.4]Dpp_4_0)([offset=-0.2]Dpp_4_6)
        \psline[linestyle=dashed]([offset=-0.2,nodesep=-1.4]Dpp_6_0)([offset=-0.2]Dpp_6_6)
        \nput[labelsep=0.7]{u}{D_0_0}{\rput[Br](-0.2,0){$\overset{j}{\rightarrow}$}}
        \nput[labelsep=0.8]{l}{D_0_0}{\rput[Br](0,0.3){$i \downarrow$}}
        \nput[labelsep=0.8]{l}{Dp_0_0}{\rput[Br](0,0.3){$i \downarrow$}}
        \nput[labelsep=0.8]{l}{Dpp_0_0}{\rput[Br](0,0.3){$i \downarrow$}}
        \nput[labelsep=0.2]{u}{D_0_8}{\rput[Bl](0.2,0){$= S$}}
        \nput[labelsep=0.2]{u}{Dp_0_7}{\rput[Bl](0.2,0){$= S'$}}
        \nput[labelsep=0.2]{u}{Dpp_0_6}{\rput[Bl](0.2,0){$= S''$}}

      \rput[bl](0,0){
        \input{DaDpaDpp.tex}
     }
    \end{pspicture}
  \end{center}
  \caption{Illustration of tables $D$, $D'$ and $D''$ with diagonal tracebacks.}
  \label{fig:DaDpaDpp}
\end{figure}

Let us consider a running example with $P = \mathtt{AGCG}$ and
$S = \mathtt{AACGGGTA}$. The values of table $D[i,j]$ are shown in the
top portion of Figure~\ref{fig:DaDpaDpp}. For example the value $D[4,3]$ is
$2$, which means that the LCSS between \texttt{AGCG} and \texttt{AAC}
has size $2$. The $D$ values can be computed with some local relations,
which we review in Section~\ref{sec-The_details}. Hence this table requires
$O(n^2)$ time to build, when $P$ and $S$ have $O(n)$ size.

The LCSSs can be recovered with tracebacks. A traceback is a pointer from a
cell $D[i,j]$ to one of its neighboring cells $D[i-1,j]$, $D[i,j-1]$ or
$D[i-1,j-1]$. The resulting paths represent the corresponding LCSSs. The
diagonal tracebacks represent matches between the corresponding strings, we
show only these tracebacks in Figure~\ref{fig:DaDpaDpp}. In our example
there is a diagonal traceback from $D[4,4]$, representing the fact that
both strings end with the letter \texttt{G}. Let us then consider
$S'= \mathtt{ACGGGTA}$ and $P' = P.\mathtt{A}$. We also need to compute the
$D$ table for these strings, shown at the in the middle of
Figure~\ref{fig:DaDpaDpp}. To avoid confusion we refer to this table as
$D'$.

We aim to compute a representation of table $D'$ in $O(n)$ time, instead of
$O(n^2)$. First let us highlight the changes between $D$ and $D'$. Column
$D[i,1]$ is removed, which corresponds to removing \texttt{A} from $S$. Row
$D'[5,j]$ is inserted, which corresponds to appending \texttt{A} to
$P$. Several values are maintained, $D[i,j+1] = D'[i,j]$. The remaining
values decrease by $1$, i.e., $D'[i,j] = D[i,j+1] - 1$. In our example only
the values of column $D'[i,0]$ decrease, the remaining values are
maintained. To determine which cells change and which remain constant we
consider another representation of table $D$. The representation used in
the Hunt-Szymanski algorithm~\cite{hunt1977fast}.

\section{Dynamic Hunt-Szymanski Algorithm}
\label{sec-The_details}
\label{sec:adapt-hunt-szym}
\begin{figure}[tb]
  \begin{center}
    \begin{pspicture}[showgrid=false](-1.5,-0.5)(8,11.0)
      {\fontfamily{cmtt}\selectfont

        \rput[Bl](1.3,10){
          \begin{psmatrix}[colsep=0.3,rowsep=0.3]
            \pscirclebox{1} & 2 & \pscirclebox{3} & \pscirclebox{4} & 5 & 6 & 7 & 8  \\
            \circlenode{Se}{A} &A & \pscirclebox{C}& \pscirclebox{G}& G& G& T& A
          \end{psmatrix}
        }

        \rput[Bl](3.0,5.5){
          \begin{psmatrix}[colsep=0.05,rowsep=0.3]
            \rnode{MA}{8}& > 2& >  \pscirclebox{1}\\
            \circlenode{MC}{3}&\\
            \rnode{MG}{6}& > 5& > \pscirclebox{4} \\
            \rnode{MT}{7}&\\
          \end{psmatrix}
        }

        \rput[Bl](2.6,8.8){\Huge $\Downarrow$}
        \rput[Bl](3.4,8.8){\psframebox[linewidth=0.05,framearc=0.6,shadow=true]{\large
            \nth{1}}}

        \rput[Bl](2.6,4.4){\Huge $\Downarrow$}
        \rput[Bl](3.4,4.4){\psframebox[linewidth=0.05,framearc=0.6,shadow=true]{\large
            \nth{2}}}

        \rput[bl](0,2){
          \rnode{F1}{8}, 2, \circlenode{C1}{1},
          \rnode{F2}{6}, 5, 4,
          \circlenode{F3}{3},
          \rnode{F4}{6}, 5, \circlenode{C3}{4}
        }
        {
          \psset{ref=b,nodesepB=-0.2,nodesepA=-0.1,rot=-90}
          \psbrace([nodesep=-0.5,offset=0.3]F2)([nodesep=-0.3,offset=0.3]F1){\rnode{FH}{A}}
          \psbrace([nodesep=-0.8,offset=0.3]F3)([nodesep=-0.3,offset=0.3]F2){G}
          \psbrace([nodesep=1.8,offset=0.3]F4)([nodesep=-0.4,offset=0.3]F4){G}
        }

        \rput(F3|[offset=0.12]FH){\rnode{L1}{C}}

        {
          \psset{linewidth=0.05,arrowsize=0.25,nodesepA=0.15,nodesepB=0.1}
          \ncline{->}{L1}{F3}
        }

        {
          \psset{shadow=true}
          \rput[Br]([offset=-2.0]C1){\circlenode{LIS1}{1}}
          \rput[Br]([offset=-2.0]F3){\circlenode{LIS2}{3}}
          \rput[Br]([offset=-2.0]C3){\circlenode{LIS3}{4}}
        }

        { \psset{linewidth=0.05,arrowsize=0.25,nodesepA=0.15,nodesepB=0.1}
          \ncline{->}{C1}{LIS1}
          \ncline{->}{F3}{LIS2}
          \ncline{->}{C3}{LIS3}
        }

        {
          \psset{shadow=true} \rput[r]([nodesep=-0.2,offset=-0.8]LIS1){A}
          \rput[r]([nodesep=-0.2,offset=-0.8]LIS2){C}
          \rput[r]([nodesep=-0.2,offset=-0.8]LIS3){G}
        }
      }

      \rput[r]([nodesep=-0.8]Se){$S = $}
      \rput[r]([nodesep=-0.4]F1|[offset=0.15]FH){$P = $}
      \rput[r]([nodesep=-0.4]F1){$L = P_S = $}
      \rput[r]([nodesep=-0.6]MA){$M_S(\mathtt{A}) = $}
      \rput[r]([nodesep=-0.6]MA|MC){$M_S(\mathtt{C}) = $}
      \rput[r]([nodesep=-0.6]MG){$M_S(\mathtt{G}) = $}
      \rput[r]([nodesep=-0.6]MT){$M_S(\mathtt{T}) = $}

    \end{pspicture}
  \end{center}
\caption{Reduction from LCSS to LIS.}
\label{fig:LisReduction}
\end{figure}
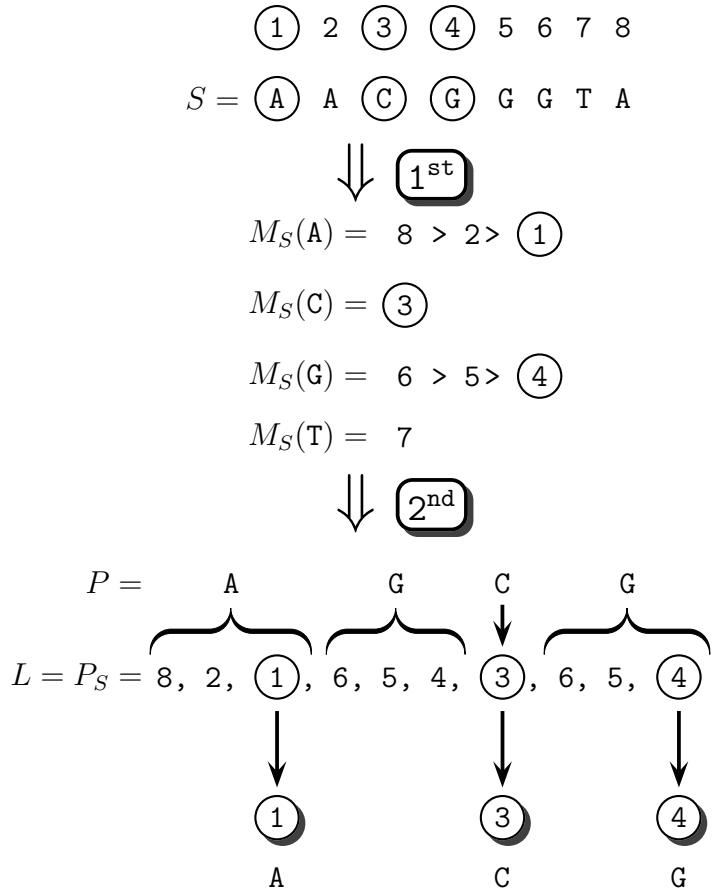

In this section we show to efficiently compute decremental string
comparison, i.e., a simple and efficient way to obtain table $D'$ from
table $D$. Let us start by reviewing and augmenting the Hunt-Szymanski
algorithm~\citep*{hunt1977fast}. The algorithm works by reducing the LCSS
to the problem of determining a longest increasing sub-sequence of numbers
(LIS). This reduction is illustrated in Figure~\ref{fig:LisReduction}. It
works in two steps. In the first step it processes $S$. For every letter
$b$ in $S$ it computes the list of positions where $b$ occurs in $S$,
represented by $M_S(b)$. In the second step it processes $P$, from left to
right, and produces a list of numbers $P_S$. For every letter $b$ of $P$
the list $M_S(b)$ is appended to the current list of numbers.

The resulting list $P_S$ consists of a list of positions of $S$, where the
same position may appear several times. Hence selecting a subsequence from
$P_S$ is equivalent to choosing letters from $S$. In our example a
resulting longest increasing subsequence is $\mathtt{1 < 3 <
  4}$. Selecting these letters from $S$ yields the desired common
subsequence \texttt{ACG}. To avoid selecting the same letter from $S$
repeatedly the LIS needs to be strictly increasing. Moreover, to guarantee
that a letter from $P$ is selected only once the lists $M_S(b)$ are sorted
in decreasing order and this order is used to build $P_S$.

\begin{figure}[tb]
  \begin{center}
    \scalebox{0.8}{
  \begin{pspicture}[showgrid=false](-4,-15.0)(6,0)
    \psset{linewidth=0.05,framearc=0.6,fillcolor=white,fillstyle=solid}
    \rput[bl](0,0){
      \psframe[shadow=true,framearc=0.2,fillcolor=black!7](-0.5,1)(7,-2.5)
    \rput[bl](0,0){
    $T_1$ :
    \rnode{S1P2}{\psframebox{8}}
      $>$
    \rnode{S1P3}{\psframebox{2}}
      $>$
    \rnode{S1P4}{\psframebox{1}}
    }
    \rput[bl](0,-1){
    $T_2$ :
    \rnode{S2P2}{\psframebox{6}}
      $>$
    \rnode{S2P3}{\psframebox{5}}
      $>$
    \rnode{S2P4}{\psframebox{4}}
      $>$
    \rnode{S2P5}{\psframebox{3}}
    }
    \rput[bl](0,-2){
    $T_3$ :
    \rnode{S3P2}{\psframebox{6}}
      $>$
    \rnode{S3P3}{\psframebox{5}}
      $>$
    \rnode{S3P4}{\psframebox{4}}
  }
}
    \rput[bl](0,-4.0){
      \psframe[shadow=true,framearc=0.2,fillcolor=black!7](-0.5,1)(7,-2.5)
    \rput[bl](0,0){
    $T_1$ :
    \rnode{S1P2}{\psframebox{8}}
      $>$
    \rnode{S1P3}{\psframebox{2}}
    }
    \rput[bl](0,-1){
    $T_2$ :
    \rnode{S2P2}{\psframebox{6}}
      $>$
    \rnode{S2P3}{\psframebox{5}}
      $>$
    \rnode{S2P4}{\psframebox{4}}
      $>$
    \rnode{S2P5}{\psframebox{3}}
    }
    \rput[bl](0,-2){
    $T_3$ :
    \rnode{S3P2}{\psframebox{6}}
      $>$
    \rnode{S3P3}{\psframebox{5}}
      $>$
    \rnode{S3P4}{\psframebox{4}}
  }
  }
    \rput[bl](0,-8.0){
      \psframe[shadow=true,framearc=0.2,fillcolor=black!7](-0.5,1)(7,-3.5)
      \rput[bl](0,0){
    $T_1$ :
    \rnode{S1P2}{\psframebox{8}}
      $>$
    \rnode{S1P3}{\psframebox{2}}
    }
    \rput[bl](0,-1){
    $T_2$ :
    \rnode{S2P2}{\psframebox{6}}
      $>$
    \rnode{S2P3}{\psframebox{5}}
      $>$
    \rnode{S2P4}{\psframebox{4}}
      $>$
    \rnode{S2P5}{\psframebox{3}}
    }
    \rput[bl](0,-2){
    $T_3$ :
    \rnode{S3P2}{\psframebox{6}}
      $>$
    \rnode{S3P3}{\psframebox{5}}
      $>$
    \rnode{S3P4}{\psframebox{4}}
    }
    \rput[bl](0,-3){
    $T_4$ :
    \rnode{S4P2}{\psframebox{8}}
    }
}
    \rput[bl](0,-13){
      \psframe[shadow=true,framearc=0.2,fillcolor=black!7](-0.5,1)(7,-2.5)
    \rput[bl](0,0){
    $T_1$ :
    \rnode{S1P2}{\psframebox{8}}
      $>$
    \rnode{S1P3}{\psframebox{6}}
      $>$
    \rnode{S1P4}{\psframebox{5}}
      $>$
    \rnode{S1P5}{\psframebox{4}}
      $>$
    \rnode{S1P6}{\psframebox{3}}
    }
    \rput[bl](0,-1){
    $T_2$ :
    \rnode{S2P2}{\psframebox{6}}
      $>$
    \rnode{S2P3}{\psframebox{5}}
      $>$
    \rnode{S2P4}{\psframebox{4}}
    }
    \rput[bl](0,-2){
    $T_3$ :
    \rnode{S3P2}{\psframebox{8}}
    }
  }
{
  \psset{arrowsize=0.5,fillstyle=none}
  \psbezier{->}(-.5,-2.0)(-2.0,-2.0)(-2.0,-4.0)(-0.5,-4.0)
  \rput[Br](-2.0,-3.0){\texttt{ExtractMin()}}
  \psbezier{->}(-.5,-6.0)(-2.0,-6.0)(-2.0,-8.0)(-0.5,-8.0)
  \rput[Br](-2.0,-6.8){\texttt{Append(8)}}
  \rput[Br](-2.0,-7.5){\texttt{Append(2)}}
  \psbezier{->}(-.5,-11.0)(-2.0,-11.0)(-2.0,-13.0)(-0.5,-13.0)
  \rput[Br](-2.0,-11.8){\texttt{ExtractMin()}}
  \rput[Br](-2.0,-12.5){\texttt{Append(8)}}
}
  \end{pspicture}
}
\end{center}
\caption{Dynamic LIS computation. The top box shows the $T_k$ lists for $P$
and $S$.}
\label{fig:LisComputation}
\end{figure}
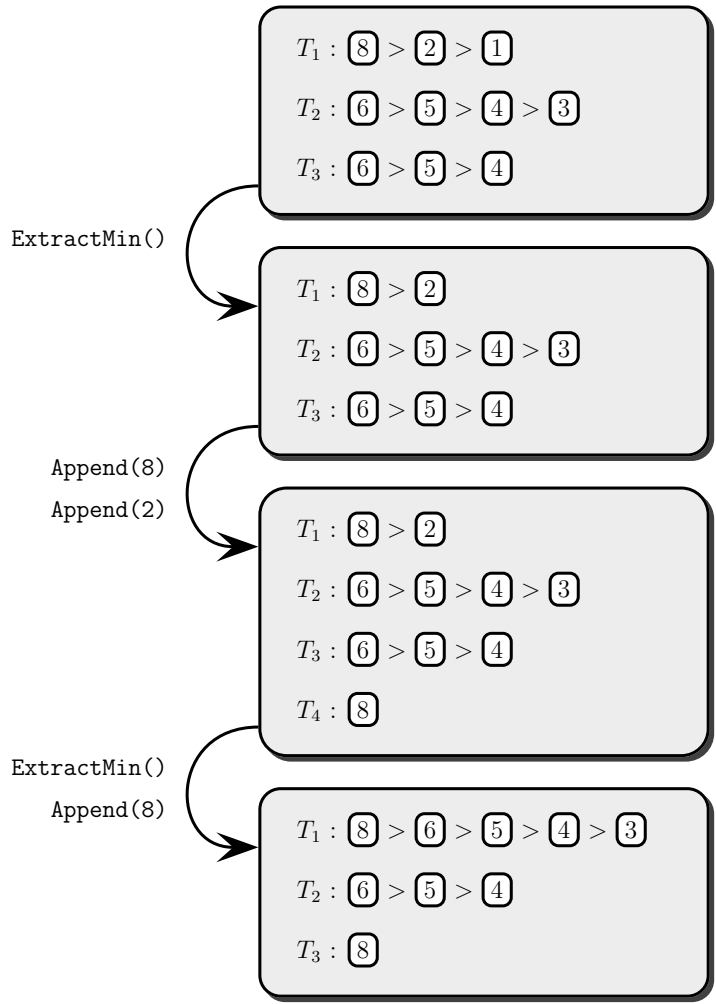

The Hunt-Szymanski algorithm then proceeds to efficiently compute the
LIS. In this context we represent the list of numbers by $L$, abstracting
away the process that was used to produce it, i.e. $L = P_S$. To determine
the LIS the algorithm uses a sequence of threshold lists $T_k$. List $T_k$
contains the value $i$ of $L$ if the longest increasing subsequence of the
first elements of $L$ up to and including $i$, has size $k$. The top box of
Figure~\ref{fig:LisComputation} shows this threshold structure for the
sequence $L$ we are considering.

Now let us return to the decremental string problem and study how this data
structure is affected when $P$ changes to $P'$ and $S$ changes to $S'$. The
top part of Figure~\ref{fig:DaDpaDpp} shows the dynamic programming table
$D$, for $P$ and $S$. The figure also illustrates $D'$ and $D''$ for the
consecutive decompositions that append letters to $P$ and remove letters
from the beginning of $S$. This figure serves to illustrate the relation
between the $D$ table and the $T_k$ lists. Figure~\ref{fig:DaDpaDpp} shows
only diagonal tracebacks, as these are the only ones that appear in the
$T_k$ lists. For example consider the cells in $D$ that are equal to $2$,
the list $T_2$ gives a representation of this set. The cells
$(2,6),(2,5),(2,4)$ and $(3,3)$ are the respective diagonal tracebacks. The
$T_k$ lists store only the $j$ coordinates, therefore the list $T_2$
contains $6 < 5 < 4 < 3$. In general list $T_k$ stores the decreasing $j$
coordinates of the cells with $D$ value $k$ and diagonal tracebacks.

Computing the threshold structure is done incrementally by processing the
elements of $L$ from left to right. Therefore the original Hunt-Szymanski
algorithm already supports the \texttt{Append} operation, which updates the
structure when a new number is appended. Each list $T_k$ is stored in
decreasing order. With this organization the sequence of tail elements is
kept in increasing order throughout the execution of the algorithm. After
processing our sample list $L$ the resulting sequence of tail elements is
$1 < 3 < 4$. This order is used to determine in which $T_k$ a given element
of $L$ should be inserted. Let us consider our running example and start
with all the $T_k$ lists empty. The first $8$ initializes $T_1$. The $2$ is
also append to $T_1$ because $8 > 2$, likewise $1$ is also appended to
$T_1$ because $2 > 1$. Number $6$ initializes list $T_2$, because $1 < 6$,
which becomes the current sequence of tail elements. Numbers $5,4$ and $3$
are also appended to $T_2$, as they are all greater than $1$ and in
decreasing order. The number $6$ initializes list $T_3$, because
$1 < 3 < 6$. Likewise $5$ and $4$ are also appended to $T_3$. Since list
$T_3$ is not empty, we know that $L$ contains an increasing subsequence of
length 3. Since $T_4$ was left empty, there is no increasing subsequence of
length $4$. Therefore the size of a LIS in our example is 3.

We now focus on adapting the threshold structure to obtain the longest
subsequence which occurs twice without overlap, in a string $F$.  Consider
$F = \mathtt{AGCGAACGGGTA}$, in which case the subsequence could be
\texttt{ACGA}, which has size 4. To compute it we can divide $F = P.S$ into
a prefix $P$ and a suffix $S$ and compute the corresponding LCSS. Notice
that this process is guaranteed to obtain a non-overlapping subsequence,
which occurs twice in $F$. However the resulting LCSS might not be the
longest such subsequence. In the division we used before the resulting
subsequence had only size $3$. Again the straightforward approach is to
compute the LCSS for all possible $P$ and $S$ pairs.

Our approach aims to avoid repeated computation by modifying the $T_k$
lists, instead of recomputing them.  We will append letters to $P$ and
delete them from to $S$. Assume that we have the $T_k$ lists for the
strings $P$ and $S$. We aim to update this structure for strings
$P' = P.\mathtt{A}$ and $S' = \mathtt{ACGGGTA}$, i.e., we want to append a
letter to $P$ and remove the first letter from $S$. Removing the first
letter from $S$ changes the $M_S(b)$ lists, in particular all the positions
are offset by $1$, for example $M_{S'}(\mathtt{G}) = \mathtt{5, 4,
  3}$. This offset does not alter the relative order of the numbers nor the
shape of the threshold structure. Therefore we ignore this offset, to
simplify the exposition. Instead assume that we start numbering the
positions of $S'$ at $2$. Now the only change is to
$M_{S'}(\mathtt{A}) = \mathtt{8, 2}$, which looses position \texttt{1}, as
it is no longer part of $S'$. In general the position that gets removed is
the overall minimum. Hence we need to apply an \texttt{ExtractMin}
operation to the threshold structure. This operation should remove all
the instances of the minimum in $L$, in this case all the instances of
$1$. In this particular example there is only one instance, but in
general there can be several such occurrences.

Due to the decreasing order of the $T_k$ lists and increasing order of the
tail elements it is straightforward to locate the overall minimum
element. The minimum is always the tail element of $T_1$. Notice that even
if there are several instances of the minimum in $L$ there is only one
element in $T_1$, because of our approach of discarding duplicated
elements. Now remove this element from $T_1$. The resulting structure still
maintains the necessary orders as the sequence of the tail elements becomes
$\mathtt{2 < 3 < 4}$, and the internal order of the $T_k$'s was not
altered. This structure is shown in the second box of
Figure~\ref{fig:LisComputation}. This is indeed the same structure that can
be obtained from the sequence $\mathtt{8,2,6,5,4,3,6,5,4}$. Thus, in this
case, no further work is required.

Since $P'$ contains an extra \texttt{A}. Hence we need to append the list
$M_{S'}(\mathtt{A})$ to $L$. Therefore we execute the operations
\texttt{Append(8)} and \texttt{Append(2)}. This alters the $T_k$ lists, as
explained above. Appending the number \texttt{8}, initializes $T_4$,
because $\mathtt{2<3<4<8}$. Appending the number $2$ does not produce any
change because it is already the tail of $T_1$. In this case we simply drop
the element, Section~\ref{sec:dynam-long-incr} describes more a elaborated
process that is used when the size of the LIS is not enough and we want to
retrieve an actual such sequence. At this point we obtained a LIS of size
$4$ which identifies a LCSS of $P'$ and $S'$ that is our goal subsequence
of $F$. However to make sure this is indeed the longest subsequence we must
continue the process and update the $T_k$ lists for $P'' = P'.\mathtt{A}$
and $S'' = \mathtt{CGGGTA}$. Again we begin by computing
\texttt{ExtractMin}. Notice that the operation removes both instances of
the number $2$. This time the operation is more elaborated because after
removing the $2$ from $T_1$ the resulting sequence of tail elements is no
longer increasing. Note that $8$ is the tail of $T_1$ and $3$ is the tail
of $T_2$ and $8 > 3$. To solve this problem we could transfer the $3$ from
$T_2$ to $T_1$, thus fixing the first inequality as $3 < 4$. However this
is not correct. Note that at this point the sequence $L$ we are considering
is $\mathtt{8,6,5,4,3,6,5,4}$, in which case $T_1$ should be
$8 > 6 > 5 > 4 > 3$. Therefore the correct procedure is to remove all the
elements from $T_2$ and append them to $T_1$. Now $T_2$ becomes empty so
all the elements from $T_3$ are moved to $T_2$, which leads $T_3$ to become
empty and therefore all the elements from $T_4$ are moved to $T_3$. Hence
$T_4$ becomes empty and the process terminates because $T_4$ was the last
list.

The general procedure for \texttt{ExtractMin} is to remove the tail element
from $T_1$ and then transfer from $T_2$ to $T_1$ all the elements that are
larger than the current tail element. The process continues from $T_{k+1}$
to $T_k$ until there are no further elements to transfer, either because
$T_{k+1}$ is empty or all its elements are larger than the current tail
element of $T_k$. In Section~\ref{sec:dynam-long-incr} we formalize, extend
and analyze this data structure. Our example finishes by appending
\texttt{8}, which gets discarded and therefore does not alter the
structure. For $P''$ and $S''$ the resulting LIS has size $3$ and is
therefore smaller than the subsequence \texttt{ACGA} obtained for $P'$ and
$S'$. This was in fact a desired subsequence, but the algorithm must scan
the remaining pairs of prefixes and suffixes to certify this conclusion.

\subsection{Implementation and Analysis}
\label{sec:dynam-long-incr}

First let us discuss which data structures can be used to efficiently store
the threshold data structure. In the classical Hunt-Szymanski algorithm
each $T_k$ list can be stored in a stack, where reading the \texttt{Top}
element and pushing new elements can be achieved in constant time. There is
no need to pop elements from the stacks, so it is enough to store the
\texttt{Top} values. These values are stored in an array so that it is
possible to perform a binary search on the top elements. The procedure to
execute \texttt{Append($i$)} is to execute a binary search on the array to
find $k$ such that \texttt{Top($T_{k-1}$)} $< i \leq$
\texttt{Top($T_{k}$)}. If $T_{k}$ is empty assume its stack top is
$+\infty$, also assume there is a sentinel list $T_0$ with
\texttt{Top($T_0$)} $= -\infty$. If for the resulting $k$ we have
\texttt{Top($T_{k}$)} $ = i$ then the procedure stops, otherwise it
performs \texttt{Push($T_k$, $i$)}.

To support the \texttt{ExtractMin} operation we prefer to use a different
data structure. We represent the $T_k$ lists using balanced binary search
trees (BST), in particular red-black trees. This allows us to compute
\texttt{Min($T_k$)}, \texttt{Insert($T_k$, $i$)}, \texttt{Remove($T_k$,
  $i$)}, \texttt{Predecessor($T_k$, $i$)}, \texttt{Split($T_k$, $v$)} and
\texttt{Concatenate($T_k$, $T_{k'}$)} in $O(\log \ell)$ time, where $\ell$
is the size of $L$. Like the Hunt-Szymanski algorithm, we keep an
array \texttt{Min[$k$]} that stores the tail element of $T_k$, so that it
can be accessed in constant time. The \texttt{Min($T_k$)} operation finds
the smallest element in $T_k$. When the BST of $T_k$ is empty it returns
$+\infty$. The \texttt{Insert($T_k$, $i$)} operation inserts the number $i$
into the BST of $T_k$. The \texttt{Remove($T_k$, $i$)} operation removes
the number $i$ from the BST of $T_k$, if key $i$ does not exist then an
error is reported and the current process is stopped. The
\texttt{Predecessor($T_k$, $i$)} operation finds the largest element of
$T_k$ that is less than or equal to $i$, i.e.,
$\max \{ j \in T_k | j \leq i \}$, the result should be a pointer to the
corresponding tree node $v$, if no such node exists the pointer should be
\texttt{NULL}. The \texttt{Split($T_k$, $v$)} operation divides the BST of
$T_k$ in two by keeping all the nodes with keys strictly larger than $v$ in
$T_k$ and putting $v$ and the remaining nodes in a new BST. The operation
\texttt{Concatenate($T_k$, $v$)} joins the BST containing node $v$ into the
BST of $T_k$, assuming that all the key values in $T_k$ are larger than or
equal to the key in $v$ and $v$ is the maximum key value in its BST. Recall
that we assume that the values in $T_k$ are not repeated, therefore the
\texttt{Insert} and \texttt{Concatenate} operations drop duplicated elements
when they occur. Algorithms~\ref{alg:append} and~\ref{alg:extractmin} show
the pseudo-code for the \texttt{Append} and \texttt{ExtractMin} operations,
respectively. Note that for the \texttt{Append} procedure the
\texttt{Min[$k$]} array plays the role of the \texttt{Top} operation in the
classical version.

Let us now analyze the time performance of the \texttt{Append} procedure,
Algorithm~\ref{alg:append}. Without the \texttt{Min[$k$]} array the overall
time would be $O((\log \ell) (\log \lambda)+\log \ell)$, where the first
term accounts for the binary search in lines 4 to 10. The second term
accounts for the \texttt{Insert} operation in line 12. Using the
\texttt{Min[$k$]} array this term reduces to $O(\log \lambda)$ and thus the
overall time becomes $O(\log \ell)$ because $\lambda \leq \ell$, since
$\lambda$ is the size of a subsequence of $L$.

Now let us analyze the \texttt{ExtractMin} procedure,
Algorithm~\ref{alg:extractmin}. The \textbf{while} loop executes at most
$\lambda$ times. Each execution requires $O(\log \ell)$ time for the
\texttt{Predecessor}, \texttt{Split} and \texttt{Concatenate}
operations. Hence we obtain a bound of at most $O(\lambda \log \ell)$
time. However an even tighter bound is possible. This operation can be
bounded by $O(\sum_{k=1}^\lambda \log(|T_k|))$, where $|T_k|$ is the size
of the list $T_k$. Because the $\log$ function is concave and the size of
all the lists adds up to $\ell$ we can use Jensen's
inequality~\citeyearpar{jensen1906} to obtain an $O(1+\log (\ell/\lambda))$
bound. The following derivation justifies the bound.
\begin{align*}
  \label{eq:1}
  \sum_{k=1}^\lambda \log(|T_k|) & = \lambda \sum_{k=1}^\lambda
                                 \frac{\log(|T_k|)}{\lambda} \\
                               & \leq \lambda \log \left( \sum_{k=1}^\lambda
                                 \frac{|T_k|}{\lambda}\right) \\
                               & = \lambda \log (\ell/\lambda)
\end{align*}

\begin{algorithm}[tb]
  \caption{\texttt{Append($i$)}}
  \label{alg:append}
  \begin{algorithmic}[1]
    \Ensure Updated threshold structure for $L$ with $i$ appended.
    \State $\ell \leftarrow (\ell + 1)$ \Comment{Increase size of $L$.}
    \State $j \leftarrow 0$
    \State $k \leftarrow (\lambda + 1)$
    \While {$j + 1< k$}
    \State $m \leftarrow \lfloor (j+k) / 2 \rfloor$
    \If {$i > $ \texttt{Min[$m$]}}
    \State $j \leftarrow m$
    \Else
    \State $k \leftarrow m$
    \EndIf
    \EndWhile
    \State \texttt{Insert($T_k$, $i$)}
    \State \texttt{Min[$k$]} $\leftarrow i$
    \If {$k = (\lambda + 1)$}
    \State $\lambda \leftarrow (\lambda + 1)$ \Comment{LIS grows}
    \EndIf
  \end{algorithmic}
\end{algorithm}

\begin{algorithm}[tb]
  \caption{\texttt{ExtractMin()}}
  \label{alg:extractmin}
  \begin{algorithmic}[1]
    \Require $L$ is not empty
    \Ensure Updated threshold structure for $L$ without the current minimum.
    \State $\ell \leftarrow (\ell - 1)$ \Comment{Decrease size of $L$.}
    \State \texttt{Remove($T_1$, Min[$1$])}
    \State $k \leftarrow 2$
    \If { $\ell > 0$ }
    \While{\texttt{Min($T_{k-1}$)} $\geq$ \texttt{Min[$k$]}}
    \State $v \leftarrow $ \texttt{Predecessor($T_k$, Min($T_{k-1}$))}
    \State \texttt{Split($T_k$, $v$)}
    \State \texttt{Concatenate($T_{k-1}$, $v$)}
    \State \texttt{Min[$k-1$]} $\leftarrow$ \texttt{Min[$k$]}
    \State $k \leftarrow k+1$
    \EndWhile
    \EndIf
    \State \texttt{Min[$k-1$]} $\leftarrow$ \texttt{Min($T_{k-1}$)}
    \If {\texttt{Min[$\lambda$]} = $+\infty$ }
    \State $\lambda \leftarrow (\lambda - 1)$ \Comment{LIS shrinks}
    \EndIf
  \end{algorithmic}
\end{algorithm}

For our particular application of the LTSS we do not need to recover an
actual sequence, at least not at the same time as identifying the $P.S$
partition of $F$. Still for a general dynamic LIS problem this may be
useful. Hence we will now explain how to augment the data structure to
support such a process.

Recall the $L$ sequences that occur in our running example. In the top of
Figure~\ref{fig:dynL} we show these sequences, numbered $L_1$, $L_2$ and
$L_3$, corresponding to the pairs of strings $(P,S)$, $(P',S')$ and
$(P'',S'')$. To retrieve the elements from the list $L$ we need to index
them. For $L_1$ this is straightforward to obtain, we simply number the
elements from $1$ to $10$. However when $L_1$ changes to $L_2$ and the
number $1$ is removed, we do not re-index the sequence. A gap is left at
position $3$. Likewise when $L_2$ changes to $L_3$ a gap is left at
position $2$. Position $12$ can be re-used because it was the last position
of $L_2$.

\begin{figure}[tb]
  \begin{center}
    \scalebox{0.8}{
    \begin{pspicture}[showgrid=false](-2.0,-6.5)(8,8.0)
      {
        \fontfamily{cmtt}\selectfont

        \rput[Bl](0.6,5.2){
          \begin{psmatrix}[colsep=0.3,rowsep=0.3]
            1& 2& 3& 4& 5& 6& 7& 8& 9& 10& 11& 12 \\
\rnode{L1}{8}& 2& 1& 6& 5& 4& 3& 6& 5& 4 \\
\rnode{L2}{8}& 2&  & 6& 5& 4& 3& 6& 5& 4 & 8 & 2\\
\rnode{L3}{8}&  &  & 6& 5& 4& 3& 6& 5& 4 & 8 & 8
          \end{psmatrix}
        }
      }
      \rput[r]([nodesep=-0.4]L1){$L_1 = $}
      \rput[r]([nodesep=-0.4]L2){$L_2 = $}
      \rput[r]([nodesep=-0.4]L3){$L_3 = $}
    \psset{linewidth=0.05,framearc=0.6,fillcolor=white,fillstyle=solid}
    \rput[bl](0,3.5){
      \psframe[shadow=true,framearc=0.2,fillcolor=black!7](-2.0,1)(8,-5.0)
      \rput[bl](0,0){
    $T_1$ :
    \rnode{S1P2}{\psframebox{8: 1}}
      $>$
    \rnode{S1P3}{\psframebox{\rnode{I}{2}: \rnode{J}{2} $<$ 12}}
    }
    \rput[bl](0,-1.5){
    $T_2$ :
    \rnode{S2P2}{\psframebox{6: 4}}
      $>$
    \rnode{S2P3}{\psframebox{5: 5}}
      $>$
    \rnode{S2P4}{\psframebox{\rnode{G}{4}: \rnode{H}{6}}}
      $>$
    \rnode{S2P5}{\psframebox{3: \rnode{HI}{7}}}
    }
    \rput[bl](0,-3.0){
    $T_3$ :
    \rnode{S3P2}{\psframebox{\rnode{C}{6}: 8}}
      $>$
    \rnode{S3P3}{\psframebox{\rnode{E}{5}: \rnode{F}{9}}}
      $>$
    \rnode{S3P4}{\psframebox{4: \rnode{D}{10}}}
    }
    \rput[bl](0,-4.5){
    $T_4$ :
    \rnode{S4P2}{\psframebox{\rnode{A}{8}: \rnode{B}{11}}}
  }

  \psset{nodesep=0.1,arrowsize=0.3,fillstyle=none}
  \ncline{->}{A}{C}
  \ncarc[arcangleA=-10,arcangleB=-40,linestyle=dashed]{->}{B}{D}
  \nbput[npos=0.6]{\texttt{Predecessor($P_3$, $11-1$)}}
  \ncarc[arcangleA=20,arcangleB=0]{->}{E}{G}
  \ncarc[arcangleA=20,arcangleB=-30,linestyle=dashed]{->}{F}{HI}
  \ncarc[arcangleA=00,arcangleB=25]{->}{G}{I}
  \naput[npos=0.85]{\texttt{Predecessor($T_1$, $4-1$)}}
  \ncarc[arcangleA=00,arcangleB=25,linestyle=dashed]{->}{H}{J}
}
        \rput[tl](1.6,-2.0){
          \psline[linewidth=0.5,linecolor=black!20,arrowsize=2,arrowlength=1]{|->}(3.8,1.8)(-0.2,1.8)

          \begin{psmatrix}[colsep=0.3,rowsep=0.2]
            $T_1$ & $T_2$ & $T_3$ & $T_4$ \\
            2:2,& 5:5,& 6:8,& 8:11\\
            2:2,& 4:6,& 6:8,& 8:11\\
            2:2,& 3:7,& 6:8,& 8:11\\
            2:2,& 4:6,& 5:9,& 8:11\\
            2:2,& 3:7,& 5:9,& 8:11\\
            2:2,& 3:7,& 4:10,& 8:11\\
          \end{psmatrix}
        }
      \end{pspicture}
}
  \end{center}
  \caption{Top: Example dynamic list $L$. Middle: Augmented threshold
    data structure for $L_2$. Bottom: Longest increasing sequences for
    $L_2$ in the order produced by Algorithm~\ref{alg:getLIS}.}
  \label{fig:dynL}
\end{figure}

To retrieve the sequences we augment the elements inside each $T_k$. Each
element stores a value of $i$ of $L$ and a list which contains positions
where $i$ occurs. These lists must contain at least one such position, but
may contain more than one. The lists contain other positions precisely to
avoid repeated elements in a $T_k$. To support LIS retrieval duplicated
elements are not dropped, instead their positions are stored in
lists. Figure~\ref{fig:dynL} shows this structure for $L_2$ where the list
$T_1$ contains the element $2$ and the position list $2, 12$. Note that
these position lists can be stored in increasing order, for each
element. Moreover the concatenation of these lists, for a fixed $T_k$, is
also sorted in increasing order. We refer to these global lists as
$P_k$. Hence for $L_2$ we have $P_1 = 1, 2, 12$ and $P_2 = 4, 5, 6, 7$ and
$P_3 = 8, 9, 10$ and $P_4 = 11$.

The possible sequences for our problem are shown in the bottom part of
Figure~\ref{fig:dynL}. Each sequence is obtained by choosing one element
from $T_1$, $T_2$, $T_3$ and $T_4$, in general one element from $T_1$ to
$T_\lambda$. The sequences must be increasing in the values chosen from
$T_k$ and also in the values chosen from $P_k$. To guarantee that searching
through these lists always yields a sequence of size $\lambda$ the
procedure starts with $k = \lambda$ and proceeds to decrease $k$. This
is illustrated by the big arrow in Figure~\ref{fig:dynL}.

In our example we start at $T_4$ and choose the value $8$ with
corresponding position $11$. Now we aim to determine which elements of
$T_3$ may occur in LIS sequences that terminate at $8$. To determine the
first such element we can compute \texttt{Predecessor($T_3$, $8-1$)}, i.e.,
find the largest value in $T_3$ that is strictly smaller than $8$. Likewise
the last such element in $P_3$ should be \texttt{Predecessor($P_3$,
  $11-1$)}, i.e., it must occur in a position strictly smaller than
$11$. Recall that $T_3$ is stored in decreasing order and $P_3$ in
increasing order. Therefore these predecessors define the interval of valid
element choices for a LIS. In general the interval of interest for a given
element $(i,p)$ of $T_{k+1}$ is between \texttt{Predecessor($T_{k}$,
  $i-1$)} and \texttt{Predecessor($P_{k}$, $p-1$)}. The
\texttt{Predecessor} on $T_k$ is obtained from the BST for $T_{k}$ in
$O(\log \ell)$ time.  The \texttt{Predecessor} on $P_k$ is conceptual and
it is enough to verify that $p' < p$, where $p'$ is the current position in
$P_{k}$. We illustrate these operations with arrows in
Figure~\ref{fig:dynL}. The dashed lines are used for $P_k$ and the filled
lines for $T_k$. The figure also illustrates the interval for element $5$
of $T_3$, i.e., \texttt{Predecessor($T_2$,$5-1$)} and
\texttt{Predecessor($P_2$,$9-1$)}. Moreover it also shows the interval for
element $4$ of $T_2$, i.e., \texttt{Predecessor($T_1$,$4-1$)} and
\texttt{Predecessor($P_1$,$6-1$)}. Therefore, iterating the
\texttt{Predecessor($T_{k}$, $i-1$)} operations, we can obtain the
lexicographically largest LIS in $O(\lambda \log \ell)$. To obtain the
remaining LIS we traverse these intervals, yielding a new LIS for each
position that is visited. Algorithm~\ref{alg:getLIS} details this
procedure.

The arguments for the \textsc{RecursiveGetLIS} are respectively, a stack
$S$, which starts empty, a value for $k$, a value $i$ of $L$ in $T_{k+1}$
and a corresponding position $p$ in $P_{k+1}$. The \texttt{Predecessor}
operation is extended to return the positions $p$, besides the $i$
values. Since the operation is on $T_k$ it returns the smallest $p$ for the
corresponding $i$, in our example \texttt{Predecessor($T_1$, $4-1$)}
returns $(2,2)$ instead of $(2,12)$. If the corresponding $T_k$ is empty
then it returns $(-\infty,+\infty)$. Moreover for this algorithm we also
use the \texttt{Next} operation, which behaves as an iterator and returns
an $(i',p')$ pair. It returns the next element, for example
\texttt{Next($P_1$)}, returns $(2,12)$, assuming it is the first invocation
after \texttt{Predecessor($T_1$, $4-1$)}. If there is no such element it
returns $(+\infty, +\infty)$. Assume that $T_k$ is represented as a BST and
$P_k$ is divided into lists, each inside a node of the BST as shown in the
middle of Figure~\ref{fig:dynL}.  The \texttt{Next} operation either moves
to the next element in the current list, or to the next node on the BST,
when it reaches the end of the current list. Note that by next on the BST
we mean a smaller value of $i$, as the $T_k$ are stored in decreasing
order. Moving to the next element on a list requires constant time, but
finding the next element on the BST may require $O(\log \ell)$ time. Hence
Algorithm~\ref{alg:getLIS} obtains each LIS in $O(\lambda \log \ell)$ time,
which again can be reduced to $O(1 + \lambda \log (\ell/\lambda))$ by
Jensen's inequality. This finishes the dynamic LIS contribution.

\begin{algorithm}[tb]
  \caption{\texttt{GetLIS()}}
  \label{alg:getLIS}
  \begin{algorithmic}[1]
    \Require $L$ is not empty
    \State \textsc{RecursiveGetLIS}($\emptyset$,$\lambda$, $+\infty$, $+\infty$)
    \Statex
    \Procedure{RecursiveGetLIS}{$S$, $k$, $i$,  $p$}
    \If{$k = 0$}
    \State \textbf{return} $S$ \Comment {Found a LIS}
    \Else
    \State $(i',p') \gets$ \texttt{Predecessor($T_{k}, i-1$)}
    \While {$p' < p$}
    \State \texttt{Push}$(S, p')$
    \State \textsc{RecursiveGetLIS}($S$,$k-1$,$i'$,$p'$)
    \State \texttt{Pop}$(S)$
    \State $(i',p') \gets$ \texttt{Next($P_{k}$)}
    \EndWhile
    \EndIf
   \EndProcedure
  \end{algorithmic}
\end{algorithm}
\begin{theorem}
\label{teo:BasicLIS}
It is possible to maintain a dynamic list with $\ell \geq 2$ numbers such
that the \texttt{Append} operation can be computed in $O(\log \ell)$ time
and \texttt{ExtractMin} and \texttt{GetLIS} requires
$O(1+\lambda \log (\ell/\lambda))$ time, for a longest increasing
sub-sequence, of size $\lambda$.
\end{theorem}
This result establishes some initial bounds of this data structure. However
these bounds are fairly non competitive for our goals. To determine an LTSS
we might generate a sequence with $\ell = O(n^2)$ elements and perform
$O(\ell)$ \texttt{Append} operations and $n$ \texttt{ExtractMin}
operations. This yields an $O(n^2 \log n)$ time algorithm. Let us improve
the performance of the dynamic LIS data structure. First we change the
red-black BSTs to finger
trees~\citep*{article,hinze_paterson_2006,guibas1978dichromatic}. This
means that \texttt{Split} and \texttt{Concatenate} operations that involve
the $t_i \geq 2$ tail elements of $T_k$ requires only $O(\log t_i)$
amortized time, instead of $O(\log |T_k|)$ time. Let us consider the
overall algorithm, from the initial empty structure to the final one. We
will analyze the overall time that is used to process a given list
$T_k$. The following argument applies for any $k$ but for simplicity
consider that we are analyzing $T_\lambda$. We have the following
inequality:
\begin{equation}
  \label{eq:5}
  \sum_{i=1}^n t_i \leq n + (\lambda - 1) n
\end{equation}
The left term in the inequality counts the number of elements that are
moved from $T_\lambda$. The right side counts the number of elements that
are removed from the data structure.
The term $n$ counts the number of elements that are
actually removed from $T_1$, one for each \texttt{ExtractMin}
operation.
The term $(\lambda-1) n$ accounts for the
elements that are dropped in the middle of the data structure. In each
\texttt{ExtractMin} operation at most $(\lambda-1)$ elements are dropped, one
for each $T_k$ list, except for $T_\lambda$.
Now the total time of these operations is
$O(\sum_{i=1}^n \log t_i)$. We can obtain the such value, restricted to
Equation~\eqref{eq:5}, by using Lagrange multipliers. We consider only one
Lagrange multiplier, represented by $c$, because we have only one
restriction. Hence the resulting Lagrangian expression is the following:
\begin{equation*}
  \sum_{i=1}^n \log t_i - c (\sum_{i=1}^n t_i - \lambda n)
\end{equation*}
A derivative in order of $t_i$ yields the following condition:
\begin{equation}
  \label{eq:6}
  \frac{1}{t_i} = c
\end{equation}
The derivative in order of $c$ returns the original restriction:
\begin{equation}
  \label{eq:7}
  \sum_{i=1}^{n} t_i = \lambda n
\end{equation}
Combining both equations we obtain that $c = 1/\lambda$ and therefore
$\log t_i = \log (\lambda)$. If we use the same upper bound for all the
other $T_k$ lists we obtain $O(n \lambda \log \lambda)$ total time for $n$
\texttt{ExtractMin} operations. This yields an amortized time of
$O(\lambda \log \lambda)$ per operation, provided the final structure is
empty. This new bound for \texttt{ExtractMin} is not necessarily smaller
than the previous $O(\lambda \log (\ell/\lambda))$, but the best of both
applies.

Besides this bound for \texttt{ExtractMin} we also need a faster
\texttt{Append} operation. Using the amortized performance of the finger
tree data structure we obtain an $O(\log \lambda)$ amortized bound for the
\texttt{Append} operation. This performance can be further improved by
discarding the binary search process. Instead do a simple linear scan from
$T_\lambda$ down to the desired position. However we do not always reset
the search, only if necessary. A sequence of decreasing numbers is
therefore refereed to as a batch. During a batch the position $k$ is not
reset. This means that processing a batch containing $m$ numbers requires
only $O(m + \lambda)$ time. Note that this is $O(1)$ amortized time per
number, when the batch contains at least $\lambda$ numbers.

A detailed description of the \texttt{AppendBatch} procedure is shown in
Algorithm~\ref{alg:appendB}, where line~\ref{line:FingerA} is computed in
$O(1)$ amortized time with the finger tree data structure and
line~\ref{line:Whilek} accumulates to $O(\lambda)$ in a decreasing
sequence. Note that the local variable $k$ preserves its value among
successive calls.
\begin{algorithm}[tb]
  \caption{\texttt{AppendBatch($i$)}}
  \label{alg:appendB}
  \begin{algorithmic}[1]
    \Ensure Updated threshold structure for $L$ with $i$ appended.  \State
    $\ell \leftarrow (\ell + 1)$ \Comment{Increase size of $L$.}  \If {
      \texttt{Min[$k$] $< i$}} \Comment{Value of $k$ is maintained between
      calls.}
    \State $k \leftarrow \lambda + 1$
    \EndIf
    \While {$k > 1$ and \texttt{Min[$T_{k-1}$] $> i$}} \label{line:Whilek}
    \State $k \leftarrow k - 1$
    \EndWhile
    \State \texttt{Insert($T_k$, $i$)} \label{line:FingerA}
    \State $\lambda \leftarrow \max(\lambda, k)$
  \end{algorithmic}
\end{algorithm}

We can now summarize our dynamic LIS data structure in the following
theorem:
\begin{theorem}
  \label{teo:LISA}
  It is possible to maintain information about the Longest Increasing
  Sub-Sequence of a dynamic list of numbers, which starts and finishes
  empty. Assuming that in total $\ell$ elements are inserted into
  the structure, in $d$ batches of decreasing sequences and also that the
  \texttt{ExtractMin} operation is executed $e$ times in total, then the
  overall time is bounded by
  $O(e \lambda (1 +\log (\min\{\lambda, \ell/\lambda \}))+ \ell + d
  \lambda)$, where $\lambda \geq 1$ is the size of the largest overall
  LIS. At anytime the size of the current LIS can be obtained in $O(1)$
  time.
\end{theorem}

We can now combine the results of Theorem~\ref{teo:BasicLIS}
and~\ref{teo:LISA} to obtain our bounds for the decremental string
comparison problem.
\begin{theorem}
  \label{teo:DEC}
  Given strings $P$ and $S$ there exists a data structure and can be used
  to obtain $\lambda \geq 2$, the size of LCSS between these strings. This
  structure requires $O(\ell)$ space, where $\ell$ is the number of matches
  between $P$ and $S$. This structure can be updated for the strings $P.c$
  and $S$ in $O(\lambda + |S|)$ time, where $c$ is any letter. It can also
  be updated for the strings $P$ and $S'$, where $S=c.S'$, in
  $O(\lambda (1 + \log(\ell/\lambda)))$ time. A sequence of operations that
  starts with an empty string and inserts letters to form the string $P$
  requires $O(|P|\lambda + \ell)$ time. A sequence of operations that
  decrements $S$ until it becomes empty requires
  $O(\min\{|S|,\ell\}\lambda (1+\log(\min \{\lambda, \ell/\lambda \})) +
  |S|)$ time.
\end{theorem}

The amortized complexities follow from Theorem~\ref{teo:LISA}, and the
extra $\min$ that appears is a bound on the number of \texttt{ExtractMin}
operations, $e$ in Theorem~\ref{teo:LISA}. This number of operations is
bounded simultaneously by $|S|$ and by $\ell$, because we cannot remove
more points than the ones that exist inside the structure. However in the
case where $e < |S|$ it is necessary to add an $O(|S|)$ term. This
corresponds to the case where the letter $c$ that is being removed from $S$
has no occurrences in $P$. In this case there is no call to
\texttt{ExtractMin} operation but this verification still needs to be
performed, which requires $O(1)$ time and must be accounted for.

Our application of computing the LTSS now follows from
Theorem~\ref{teo:DEC}. The total amount of time the LTSS algorithm is
therefore
$O(\min\{n,\ell\} \lambda (1+\log (\min\{\lambda, \ell/\lambda\})) + n +
\ell)$, where $\lambda \geq 2$ is the size of the LTSS and $\ell \geq 2$ is
the number of pairs of positions in $F$ that contain the same letter.

\section{Related work} 
\label{sec-Related_work}
An initial efficient algorithm to compute the LTSS, for the simple case of
only one string $F$, was given by~\citet*{kosowski2004efficient}. This
algorithm required optimal $O(n^2)$ time and $O(n)$ space. \citet*[Section
5.6]{DBLP:journals/mics/Tiskin08} presented an algorithm which obtains the
smallest worst case bound by exploring the Monge properties of the
respective distance matrices. This property depends on the fact
that the graph underlying the $D$ table of two strings is planar.  The
resulting algorithm obtains the overall worst case time bound
of~$O(n^2 (\log \log n)^2/(\log n)^2)$.

The work on incremental string comparison was initiated
by~\citet*{landau1998incremental}, which obtained an $O(n)$ time algorithm
to obtain $D'$ from $D$. A simpler version, with the same performance was
presented by~\citet*{kim2000dynamic}, which is simultaneously incremental
and decremental. This is the first instance of the decremental variation of
the problem. This solution was presented for the edit
distance.~\citet*{ishida2005fully} presented an algorithm which reduced the
time complexity from $O(n)$ to $O(\lambda)$ and was fully incremental. The
algorithm was presented for the LCSS and they also reduced the space
requirements from $O(n^2)$ to $O(n \lambda)$.

\citet*{landau2004two} studied the problem of consecutive suffix
alignment problem, which obtained the size of the LCSS between all the
suffixes of a string $A$ and a string $B$, the final version of the paper
appeared in~\citeyearpar{LANDAU20071095}. The authors presented two
algorithms for this problem, which required $O(n \lambda)$ and
$O(n \lambda + n \log \sigma)$ time, where $\sigma$ is the size of the
alphabet of the underlying strings. Their approach uses a structure similar
to the $T_k$ lists from the Hunt-Szymanski algorithm, but contrary to our
approach of Section~\ref{sec:adapt-hunt-szym} the elements are prepended to
a variation of the $T_k$ lists. Moreover their structure is not
decremental. Because of these nuances the relation to LTSS is not immediate
which justifies the algorithm of~\citet*{kosowski2004efficient}, in the
same year.

A corner stone of all these results is the algorithm
from~\citet*{hunt1977fast}, whose crucial idea was the reduction from the

LCSS to the LIS, although this was not immediately clear in the original
presentation. It was partially identified
by~\cite{apostolico1986improving},~\cite{apostolico1987longest} and made
explicit by~\cite{jacobson1992heaviest} and independently
by~\cite{pevzner1992matrix}. Interestingly the original presentation
of~\citet{hunt1977fast} reported an $O((n+ \ell) \log \ell)$ time bound,
where $\ell$ is the size of the sequence $L$. This is a significant
improvement over the plain dynamic programming algorithm, which always
requires $O(n^2)$ time. Although in the worst case $\ell$ may be $O(n^2)$,
in general it may be significantly smaller.  The original
complexity was not always faster than the plain algorithm, because $\ell$
may be $\Omega(n / \log n)$. This issue was addressed
by~\citet{apostolico1986improving} which obtained $O(n^2)$ time worst case
guarantees.  Their algorithm already considered using finger
trees to represent the $T_k$ lists. Improvements of the Hunt-Szymanski
algorithm based on bitwise operations where proposed
by~\citet*{crochemore2003speeding}.

A data structure that supports dynamic longest increasing sub-sequences was
presented by~\citet*{chen2013dynamic}. The focus is in supporting
insertions anywhere in the sequence, which is achieved in
$O(1 + \lambda \log (\ell/\lambda))$ time. The authors obtain one
corresponding LIS in $O(\lambda + \log \ell)$ time. This is more efficient
than the procedure we explain before Theorem~\ref{teo:BasicLIS}, however
our procedure can be used to obtain all the sub-sequences, whereas their
approach obtains only one. Their data structure is similar to the one we
present, which is expected as both are related to the Hunt-Szymanski
algorithm. \citet{chen2013dynamic} use level key lists $L_k$, which are
similar to our $T_k$ lists, but store index value pairs and are sorted by
increasing index. This is similar to the structure we use for the
\texttt{GetLIS} operation, but the lists are flattened, instead of storing
the indexes in a second structure. Moreover they also use red-black trees
to split and concatenate lists and also mention exploring fingering
properties of the structure. The presentation mentions deletions but the
focus is on insertions. It seems plausible their representation could also
support deletions efficiently.

The most recent approach for computing the LTSS was proposed
by~\citet*{inoue2020longest}. Their algorithm is very similar to the one we
present in this paper. They also reduce the problem to a dynamic LIS
problem and used the data structure of~\citet*{chen2013dynamic} to obtain a
complexity of
$O(\min\{n,\ell\} \lambda (1+\log ( \ell/\lambda)) + n + \ell \log n)$. In
the next section we explain how our algorithm improves upon their result
and discuss future possible improvements.

\section{Conclusions} 
\label{sec-Conclusions_and_further_work}

In this section we recall and discuss the contributions of the paper in
context. In this paper we presented a new algorithm to determine the
longest tandem scattered sub-sequence of a string $F$. In the process we
introduced the decremental string comparison problem and provided new data
structures to support dynamic LIS sequences. We studied a dynamic version
of the Hunt-Szymanski algorithm, which yielded several interesting
results. Considering the LTSS problem itself the strongest work case bounds
where obtained by~\citet*{DBLP:journals/mics/Tiskin08} with an
$O(n^2 (\log \log n)^2/(\log n)^2)$ time bound. Both this algorithm and one
by~\citet*{kosowski2004efficient} seem to have the average case with the
same bound as the worst case. The algorithm we obtain as a consequence of
Theorem~\ref{teo:LISA} obtains
$O(\min\{n,\ell\} \lambda (1+\log (\min\{\lambda, \ell/\lambda\})) + n +
\ell)$ time, where $\lambda \geq 2$ is the size of the LTSS and
$\ell \geq 2$ is the number of pairs of positions in $F$ that contain the
same letter. Hence when $\ell = o(n^2 (\log \log n)^2/(\log n)^2)$ and
$\lambda =o(n (\log \log n)^2/((\log (\min \{\lambda,\ell/\lambda\})) (\log
n)^2))$ our algorithm becomes more efficient. Hence our algorithm is most
efficient when the size of the LTSS is small, the extreme case in favor of
our algorithm occurs when all the letters in $F$ are distinct. In this case
our algorithm is actually linear, i.e., $O(n)$ time and space. This
particular case is trivial but a similar situation occurs when the alphabet
size is large, i.e., polylog. This is also the case were the original
Hunt-Szymanski algorithm obtains its best performance.

One important contribution of this work is the relation between the LTSS
and the decremental and incremental string comparison algorithms. This
relation is straightforward but seems to have remained
unnoticed\footnote{It was also recently pointed out
  by~\citet*{inoue2020longest}.}, as the algorithm
of~\citet{ishida2005fully} also depends on $\lambda$ and could thus be used
to compute the LTSS, if the structure was also decremental. On the other
hand the structure of~\citet{kim2000dynamic} is decremental but does not
depend on $\lambda$. This relation was indeed explored in the work
of~\citet*{DBLP:journals/mics/Tiskin08}, but as mentioned above the
resulting algorithm is also not dependent on $\lambda$. Hence for the case
of a single string the algorithm we presented in
Section~\ref{sec:adapt-hunt-szym} yields competitive results. In fact it is
very interesting to compare the $T_k$ lists of the Hunt-Szymanski algorithm
to the incremental data structure of~\citet{ishida2005fully}. In essence
their structure consists in expanded $T_k$ lists, where each element is
repeated several times so that the list becomes size $n$. This increases
the space requirements but makes navigating the lists and across lists more
convenient. Also it forces one of the operations be $O(n)$ and thus the
overall bound is always $O(n^2)$ instead of $O(n \lambda)$.

The recent work by~\citet*{inoue2020longest} follows essentially the same
approach as this paper. It solves the LTSS problem by resorting to the same
decremental string comparison approach and solves this problem using
the Hunt-Szymanski reduction to a LIS problem. A dynamic version of the LIS
problem that supports the \texttt{ExtractMin} operation is also
considered. In fact we were unaware of the similarity of their approach
until recently. Still our approach contains several key insights which
allow us to obtain a result that is competitive against their
$O(\min\{n,\ell\} \lambda (1+\log (\ell/\lambda)) + n + \ell \log n)$ time
bound\footnote{Note that we added an $O(n)$ term to their complexity
  result, because in the case that we considered when all the letters of
  $F$ are distinct we have $\ell = \lambda = 0$, but their algorithm still
  requires $O(n)$ time, as does ours.}. They use essentially the dynamic
LIS structure of~\citet*{chen2013dynamic} and propose only one improvement,
batched \texttt{ExtractMin} operations. This means that they obtain better
performance for a sequence of \texttt{ExtractMin} operations. We obtain the
same improvement by using our duplicate discarding approach. This means
that a single \texttt{ExtractMin} operation on our data structure
corresponds to several on theirs, because their \texttt{ExtractMin}
operation removes one duplicate at a time whereas ours removes all. Hence
our \texttt{ExtractMin} operation requires the same time as their batch of
\texttt{ExtractMin}. One very important optimization of our approach are
the finger trees to represent the $T_k$ lists which lead to the improved
performance of the \texttt{AppendBatch} operation,
Algorithm~\ref{alg:appendB}. This implies that there is no $O(\log n)$
factor associated to the $\ell$ term in our complexity. This term is most
likely to dominate the overall complexity in several interesting cases and
in our algorithm it is $O(\ell)$ whereas in theirs it is $O(\ell \log
n)$. However this is a tradeoff as we obtain an extra $O(n \lambda)$ term,
whereas theirs is only $O(n)$. Hence ignoring the first term the resulting
comparison is between $O(n + \ell \log n)$ for their algorithm and
$O(n \lambda + \ell)$ for ours. Hence our algorithm obtains better
performance provided that $\ell = \Omega(n (\lambda -1)/((\log n)-1))$ and
$n \leq \ell$ because of the first term. Let us consider a very simple
example where this is likely to happen. Assume that the letters of $F$ are
obtained independently and uniformly at random from an alphabet of size
$\sigma$. In this case $\ell$ is expected to be $(n/\sigma)^2$ and the
distribution is highly concentrated around this value. Hence in order for
our algorithm to obtain the best theoretical bound it is necessary to have
an alphabet size $\sigma$ that is smaller than
$\sqrt{(n((\log n)-1))/(\lambda-1)}$. This is actually a very loose bound,
much larger than poly logarithmic alphabets. Hence our algorithm's
theoretical bound yields the best performance for most alphabets, except
for exceedingly large ones. Even in extremely large alphabets there is the
mitigating expectation that $\sigma$ and $\lambda$ have an inverse
relation, meaning that larger values of $\sigma$ should yield smaller
values of $\lambda$.

The final improvement on the work of~\citet*{inoue2020longest} is the
analysis with Lagrange multipliers that yields the
$\log(\min\{\lambda, \ell/\lambda\})$ bound that improves on the previous
$\log(\ell/\lambda)$ complexity. Also we believe that future research on
these data structures will focus precisely on this factor. One approach
that seems promising is to store the $T_k$ lists in a data structure that
supports the dynamic fractional cascading technique
of~\citet*{Chazelle1986}, potentially reducing this factor to
$O(\log \log n)$.

The final contribution of this paper is the data structure to maintain a
dynamic LIS. Our approach uses a couple of nuances that allow us to obtain
Theorem~\ref{teo:LISA}. Using the \texttt{AppendBatch} of
Algorithm~\ref{alg:appendB} the complexity of the original Hunt-Szymanski
algorithm drops to $O(\ell + n \lambda)$, which is never more than $O(n^2)$
and sometimes much better. Given the importance of this algorithm, similar
improvements have already been proposed
by~\cite{apostolico1986improving}. Still our work provides a fairly simple
alternative.

\section*{Acknowledgements}
\label{sec:acknowledgements}
We are grateful to Hideo Bannai, Travis Gagie, Gary Hoppenworth, Simon
J. Puglisi and Tatiana Rocher, for interesting discussions on this topic,
at StringMasters, Lisbon 2018. We dedicate a special thanks to Hideo for
suggesting this problem and providing insightful comments on a preliminary
draft of this paper.

The work reported in this article was supported by national funds through
Funda\c{c}\~ao para a Ci\^encia e Tecnologia (FCT) with reference
UIDB/50021/2020 and through project NGPHYLO PTDC/CCI-BIO/29676/2017. Funded
in part by European Union’s Horizon 2020 research and innovation programme
under the Marie Sk{\l}odowska-Curie Actions grant agreement No 690941.

 \bibliographystyle{elsarticle-harv}
 \bibliography{elsarticle-template-harv}

\end{document}

%% file: DaDpaDpp.tex
  \begin{psmatrix}[colsep=0.4,rowsep=0.1]
    & [name=D_0_0] $0$& [name=D_0_1] $0$& [name=D_0_2] $0$& [name=D_0_3] $0$& [name=D_0_4] $0$& [name=D_0_5] $0$& [name=D_0_6] $0$& [name=D_0_7] $0$& [name=D_0_8] $0$\\
    & [name=D_1_0] $0$& [name=D_1_1] $1$& [name=D_1_2] $1$& [name=D_1_3] $1$& [name=D_1_4] $1$& [name=D_1_5] $1$& [name=D_1_6] $1$& [name=D_1_7] $1$& [name=D_1_8] $1$\\
    & [name=D_2_0] $0$& [name=D_2_1] $1$& [name=D_2_2] $1$& [name=D_2_3] $1$& [name=D_2_4] $2$& [name=D_2_5] $2$& [name=D_2_6] $2$& [name=D_2_7] $2$& [name=D_2_8] $2$\\
    & [name=D_3_0] $0$& [name=D_3_1] $1$& [name=D_3_2] $1$& [name=D_3_3] $2$& [name=D_3_4] $2$& [name=D_3_5] $2$& [name=D_3_6] $2$& [name=D_3_7] $2$& [name=D_3_8] $2$\\
    & [name=D_4_0] $0$& [name=D_4_1] $1$& [name=D_4_2] $1$& [name=D_4_3] $2$& [name=D_4_4] $3$& [name=D_4_5] $3$& [name=D_4_6] $3$& [name=D_4_7] $3$& [name=D_4_8] $3$\\
    \\
    & & [name=Dp_0_0] $0$& [name=Dp_0_1] $0$& [name=Dp_0_2] $0$& [name=Dp_0_3] $0$& [name=Dp_0_4] $0$& [name=Dp_0_5] $0$& [name=Dp_0_6] $0$& [name=Dp_0_7] $0$\\
    & & [name=Dp_1_0] $0$& [name=Dp_1_1] $1$& [name=Dp_1_2] $1$& [name=Dp_1_3] $1$& [name=Dp_1_4] $1$& [name=Dp_1_5] $1$& [name=Dp_1_6] $1$& [name=Dp_1_7] $1$\\
    & & [name=Dp_2_0] $0$& [name=Dp_2_1] $1$& [name=Dp_2_2] $1$& [name=Dp_2_3] $2$& [name=Dp_2_4] $2$& [name=Dp_2_5] $2$& [name=Dp_2_6] $2$& [name=Dp_2_7] $2$\\
    & & [name=Dp_3_0] $0$& [name=Dp_3_1] $1$& [name=Dp_3_2] $2$& [name=Dp_3_3] $2$& [name=Dp_3_4] $2$& [name=Dp_3_5] $2$& [name=Dp_3_6] $2$& [name=Dp_3_7] $2$\\
    & & [name=Dp_4_0] $0$& [name=Dp_4_1] $1$& [name=Dp_4_2] $2$& [name=Dp_4_3] $3$& [name=Dp_4_4] $3$& [name=Dp_4_5] $3$& [name=Dp_4_6] $3$& [name=Dp_4_7] $3$\\
    & & [name=Dp_5_0] $0$& [name=Dp_5_1] $1$& [name=Dp_5_2] $2$& [name=Dp_5_3] $3$& [name=Dp_5_4] $3$& [name=Dp_5_5] $3$& [name=Dp_5_6] $3$& [name=Dp_5_7] $4$\\
    \\
    & & & [name=Dpp_0_0] $0$& [name=Dpp_0_1] $0$& [name=Dpp_0_2] $0$& [name=Dpp_0_3] $0$& [name=Dpp_0_4] $0$& [name=Dpp_0_5] $0$& [name=Dpp_0_6] $0$\\
    & & & [name=Dpp_1_0] $0$& [name=Dpp_1_1] $0$& [name=Dpp_1_2] $0$& [name=Dpp_1_3] $0$& [name=Dpp_1_4] $0$& [name=Dpp_1_5] $0$& [name=Dpp_1_6] $1$\\
    & & & [name=Dpp_2_0] $0$& [name=Dpp_2_1] $0$& [name=Dpp_2_2] $1$& [name=Dpp_2_3] $1$& [name=Dpp_2_4] $1$& [name=Dpp_2_5] $1$& [name=Dpp_2_6] $1$\\
    & & & [name=Dpp_3_0] $0$& [name=Dpp_3_1] $1$& [name=Dpp_3_2] $1$& [name=Dpp_3_3] $1$& [name=Dpp_3_4] $1$& [name=Dpp_3_5] $1$& [name=Dpp_3_6] $1$\\
    & & & [name=Dpp_4_0] $0$& [name=Dpp_4_1] $1$& [name=Dpp_4_2] $2$& [name=Dpp_4_3] $2$& [name=Dpp_4_4] $2$& [name=Dpp_4_5] $2$& [name=Dpp_4_6] $2$\\
    & & & [name=Dpp_5_0] $0$& [name=Dpp_5_1] $1$& [name=Dpp_5_2] $2$& [name=Dpp_5_3] $2$& [name=Dpp_5_4] $2$& [name=Dpp_5_5] $2$& [name=Dpp_5_6] $3$\\
    & & & [name=Dpp_6_0] $0$& [name=Dpp_6_1] $1$& [name=Dpp_6_2] $2$& [name=Dpp_6_3] $2$& [name=Dpp_6_4] $2$& [name=Dpp_6_5] $2$& [name=Dpp_6_6] $3$\\
  \end{psmatrix}
  \nput[labelsep=0.2]{u}{D_0_1}{\texttt{A}}
  \nput[labelsep=0.2]{u}{D_0_2}{\texttt{A}}
  \nput[labelsep=0.2]{u}{D_0_3}{\texttt{C}}
  \nput[labelsep=0.2]{u}{D_0_4}{\texttt{G}}
  \nput[labelsep=0.2]{u}{D_0_5}{\texttt{G}}
  \nput[labelsep=0.2]{u}{D_0_6}{\texttt{G}}
  \nput[labelsep=0.2]{u}{D_0_7}{\texttt{T}}
  \nput[labelsep=0.2]{u}{D_0_8}{\texttt{A}}
  \nput[labelsep=0.7]{u}{D_0_0}{$0$}
  \nput[labelsep=0.7]{u}{D_0_1}{$1$}
  \nput[labelsep=0.7]{u}{D_0_2}{$2$}
  \nput[labelsep=0.7]{u}{D_0_3}{$3$}
  \nput[labelsep=0.7]{u}{D_0_4}{$4$}
  \nput[labelsep=0.7]{u}{D_0_5}{$5$}
  \nput[labelsep=0.7]{u}{D_0_6}{$6$}
  \nput[labelsep=0.7]{u}{D_0_7}{$7$}
  \nput[labelsep=0.7]{u}{D_0_8}{$8$}
  \nput[labelsep=0.31]{150}{D_0_0}{$P$}
  \nput[labelsep=0.4]{l}{D_1_0}{\texttt{A}}
  \nput[labelsep=0.4]{l}{D_2_0}{\texttt{G}}
  \nput[labelsep=0.4]{l}{D_3_0}{\texttt{C}}
  \nput[labelsep=0.4]{l}{D_4_0}{\texttt{G}}
  \nput[labelsep=0.8]{l}{D_0_0}{$0$}
  \nput[labelsep=0.8]{l}{D_1_0}{$1$}
  \nput[labelsep=0.8]{l}{D_2_0}{$2$}
  \nput[labelsep=0.8]{l}{D_3_0}{$3$}
  \nput[labelsep=0.8]{l}{D_4_0}{$4$}
  \ncline{->}{D_1_1}{D_0_0}
  \ncline{->}{D_1_2}{D_0_1}
  \ncline{->}{D_1_8}{D_0_7}
  \ncline{->}{D_2_4}{D_1_3}
  \ncline{->}{D_2_5}{D_1_4}
  \ncline{->}{D_2_6}{D_1_5}
  \ncline{->}{D_3_3}{D_2_2}
  \ncline{->}{D_4_4}{D_3_3}
  \ncline{->}{D_4_5}{D_3_4}
  \ncline{->}{D_4_6}{D_3_5}
  \nput[labelsep=0.2]{u}{Dp_0_1}{\texttt{A}}
  \nput[labelsep=0.2]{u}{Dp_0_2}{\texttt{C}}
  \nput[labelsep=0.2]{u}{Dp_0_3}{\texttt{G}}
  \nput[labelsep=0.2]{u}{Dp_0_4}{\texttt{G}}
  \nput[labelsep=0.2]{u}{Dp_0_5}{\texttt{G}}
  \nput[labelsep=0.2]{u}{Dp_0_6}{\texttt{T}}
  \nput[labelsep=0.2]{u}{Dp_0_7}{\texttt{A}}
  \nput[labelsep=0.28]{150}{Dp_0_0}{$P'$}
  \nput[labelsep=0.4]{l}{Dp_1_0}{\texttt{A}}
  \nput[labelsep=0.4]{l}{Dp_2_0}{\texttt{G}}
  \nput[labelsep=0.4]{l}{Dp_3_0}{\texttt{C}}
  \nput[labelsep=0.4]{l}{Dp_4_0}{\texttt{G}}
  \nput[labelsep=0.4]{l}{Dp_5_0}{\texttt{A}}
  \nput[labelsep=0.8]{l}{Dp_0_0}{$0$}
  \nput[labelsep=0.8]{l}{Dp_1_0}{$1$}
  \nput[labelsep=0.8]{l}{Dp_2_0}{$2$}
  \nput[labelsep=0.8]{l}{Dp_3_0}{$3$}
  \nput[labelsep=0.8]{l}{Dp_4_0}{$4$}
  \nput[labelsep=0.8]{l}{Dp_5_0}{$5$}
  \ncline{->}{Dp_1_1}{Dp_0_0}
  \ncline{->}{Dp_1_7}{Dp_0_6}
  \ncline{->}{Dp_2_3}{Dp_1_2}
  \ncline{->}{Dp_2_4}{Dp_1_3}
  \ncline{->}{Dp_2_5}{Dp_1_4}
  \ncline{->}{Dp_3_2}{Dp_2_1}
  \ncline{->}{Dp_4_3}{Dp_3_2}
  \ncline{->}{Dp_4_4}{Dp_3_3}
  \ncline{->}{Dp_4_5}{Dp_3_4}
  \ncline{->}{Dp_5_1}{Dp_4_0}
  \ncline{->}{Dp_5_7}{Dp_4_6}
  \nput[labelsep=0.2]{u}{Dpp_0_1}{\texttt{C}}
  \nput[labelsep=0.2]{u}{Dpp_0_2}{\texttt{G}}
  \nput[labelsep=0.2]{u}{Dpp_0_3}{\texttt{G}}
  \nput[labelsep=0.2]{u}{Dpp_0_4}{\texttt{G}}
  \nput[labelsep=0.2]{u}{Dpp_0_5}{\texttt{T}}
  \nput[labelsep=0.2]{u}{Dpp_0_6}{\texttt{A}}
  \nput[labelsep=0.25]{150}{Dpp_0_0}{$P''$}
  \nput[labelsep=0.4]{l}{Dpp_1_0}{\texttt{A}}
  \nput[labelsep=0.4]{l}{Dpp_2_0}{\texttt{G}}
  \nput[labelsep=0.4]{l}{Dpp_3_0}{\texttt{C}}
  \nput[labelsep=0.4]{l}{Dpp_4_0}{\texttt{G}}
  \nput[labelsep=0.4]{l}{Dpp_5_0}{\texttt{A}}
  \nput[labelsep=0.4]{l}{Dpp_6_0}{\texttt{A}}
  \nput[labelsep=0.8]{l}{Dpp_0_0}{$0$}
  \nput[labelsep=0.8]{l}{Dpp_1_0}{$1$}
  \nput[labelsep=0.8]{l}{Dpp_2_0}{$2$}
  \nput[labelsep=0.8]{l}{Dpp_3_0}{$3$}
  \nput[labelsep=0.8]{l}{Dpp_4_0}{$4$}
  \nput[labelsep=0.8]{l}{Dpp_5_0}{$5$}
  \nput[labelsep=0.8]{l}{Dpp_6_0}{$6$}
  \ncline{->}{Dpp_1_6}{Dpp_0_5}
  \ncline{->}{Dpp_2_2}{Dpp_1_1}
  \ncline{->}{Dpp_2_3}{Dpp_1_2}
  \ncline{->}{Dpp_2_4}{Dpp_1_3}
  \ncline{->}{Dpp_3_1}{Dpp_2_0}
  \ncline{->}{Dpp_4_2}{Dpp_3_1}
  \ncline{->}{Dpp_4_3}{Dpp_3_2}
  \ncline{->}{Dpp_4_4}{Dpp_3_3}
  \ncline{->}{Dpp_5_6}{Dpp_4_5}
  \ncline{->}{Dpp_6_6}{Dpp_5_5}